\documentclass[aps,prb,twocolumn,showpacs,superscriptaddress]{revtex4-1}
\usepackage{graphicx}

\begin{document}

\title{Neutron scattering study of correlated phase behavior in Sr$_2$IrO$_4$}

\author{Chetan Dhital}
\affiliation{Department of Physics, Boston College, Chestnut Hill, Massachusetts 02467, USA}
\author{Tom Hogan}
\affiliation{Department of Physics, Boston College, Chestnut Hill, Massachusetts 02467, USA}
\author{Z. Yamani}
\affiliation{Canadian Neutron Beam Centre, National Research Council, Chalk River, Ontario, Canada K0J 1P0}
\author{Clarina de la Cruz}
\affiliation{Neutron Scattering Science Division, Oak Ridge National Laboratory, Oak Ridge, Tennessee 37831-6393, USA}
\author{Xiang Chen}
\affiliation{Department of Physics, Boston College, Chestnut Hill, Massachusetts 02467, USA}
\author{Sovit Khadka}
\affiliation{Department of Physics, Boston College, Chestnut Hill, Massachusetts 02467, USA}
\author{Zhensong Ren}
\affiliation{Department of Physics, Boston College, Chestnut Hill, Massachusetts 02467, USA}
\author{Stephen D. Wilson}
\email{stephen.wilson@bc.edu}
\affiliation{Department of Physics, Boston College, Chestnut Hill, Massachusetts 02467, USA}

\begin{abstract}
Neutron diffraction measurements are presented exploring the magnetic and structural phase behaviors of the candidate J$_{eff}=1/2$ Mott insulating iridate Sr$_2$IrO$_4$. Comparisons are drawn between the correlated magnetism in this single layer system and its bilayer analog Sr$_3$Ir$_2$O$_7$ where both materials exhibit magnetic domains originating from crystallographic twinning and comparable moment sizes. Weakly temperature dependent superlattice peaks violating the reported tetragonal space group of Sr$_2$IrO$_4$ are observed supporting the notion of a lower structural symmetry arising from a high temperature lattice distortion, and we use this to argue that moments orient along a unique in-plane axis demonstrating an orthorhombic symmetry in the resulting spin structure.  Our results demonstrate that the correlated spin order and structural phase behaviors in both single and bilayer Sr$_{n+1}$Ir$_{n}$O$_{3n+1}$ systems are remarkably similar and suggest comparable correlation strengths in each system.                    
\end{abstract}


\maketitle
\section{Introduction}
The electronic ground states and the relative role of spin orbit coupling and correlation effects within several classes of perovskite iridates have recently been a subject of debate \cite{kimscience, gedhik, kee, kim214}.  Specifically, within the Ruddelsden-Popper (RP) Sr$_{2n+1}$Ir$_n$O$_{3n+1}$ series, a picture of $J_{eff}=1/2$ Mott insulating phase has been proposed to explain the antiferromagnetically ordered, insulating ground states of the $n=1$ (Sr$_{2}$IrO$_{4}$) and $n=2$ (Sr$_3$Ir$_2$O$_7$) members of the iridate RP series \cite{kim214, kim327, clancy, clancyXAFS}.  Within this picture, a cooperative interplay of spin-orbit induced bandwidth narrowing and on-site Coulomb interactions generates the necessary Mott insulating ground state.   An alternate interpretation, however, has also been proposed that instead models the insulating ground states of these systems as arising from a weakly correlated band insulator in which the formation of magnetic order continuously builds the bandgap and where the effects of electron-electron correlations are secondary \cite{arita,kee}.  

The insulating ground state in the Sr$_{2n+1}$Ir$_n$O$_{3n+1}$ series is destabilized as the dimensionality is increased with increasing $n$;\cite{moon} however the relative changes in the electronic and magnetic properties as the materials transition from the $n=1$ single layer Sr$_2$IrO$_4$ (Sr-214) to the $n=2$ bilayer Sr$_3$Ir$_2$O$_7$ (Sr-327) remain poorly understood.  The room temperature resistivity of Sr-214 and Sr-327 reflect the notion of a reduced electronic bandgap with increased dimensionality \cite{cao214, cao327}; however their low temperature properties differ substantially \cite{cao327, dhital, cao214}.  While charge transport in Sr-214 does not directly couple to the onset of magnetic order, Arrhenius activated transport appears below 230K, which then transitions to a variable range hopping (VRH) regime at low temperatures \cite{kini}.  In contrast to this, no simple model for the low temperature transport of Sr-327 applies\cite{dhital, nagai}. Instead the resistivity couples to the onset of antiferromagnetic (AF) order below 280K via an enhanced spin-charge coupling mechanism and a proposed second phase transition appears below $T^{*}\approx70$ K\cite{cao327, dhital}.  

While both systems are canted G-type antiferromagnets,\cite{kimscience,fujiyama214,bossegia} the moments in Sr-214 orient within the basal plane whereas those for Sr-327 orient predominantly along the c-axis\cite{kim327}.  The driving force behind this reorientation stems from the enhanced interplane coupling of the bilayer compound; however the evolution of the effective ordered moment and magnetism as a function of increasing dimensionality remains largely unexplored. Historical comparisons in the literature rely on saturated net moments as observed via bulk magnetization\cite{kimscience, bossegia, cao214, cao327, fujiyama}. This is very imprecise given that these materials are canted antiferromagnets with entirely different moment orientations and hence vastly different pictures of ferromagnetic moments generated via in-plane canting.   Given that the relative ordered moment size constitutes one traditional metric for ascertaining the relative strength of correlations between two materials in similar electronic environments, a precise determination of the relative values of their ordered moments is particularly germane.

In this paper, we determine the AF ordered moment in the Sr-214 system, allowing for a direct comparison with the bilayer Sr-327 system using the conventionally accepted models of spin order in each material.  Our neutron scattering measurements reveal that both materials possess near identical magnetic order parameters with comparable ordered moments despite the picture of the near metallic phase of Sr-327.  In addition, within Sr-214 we resolve the presence of high-temperature superlattice peaks at positions forbidden by the nominal $I4_1/acd$ tetragonal space group as well as the presence of magnetic domains consistent with an orthorhombic structural symmetry.  This high-temperature superlattice likely arises from unreported oxygen distortions within the lattice relaxing through a high temperature structural phase transition.  Our results point toward surprisingly similar pictures of correlated electronic phase behavior in the Sr-214 and Sr-327 systems despite the proposed picture of the marginally insulating state of Sr-327.

\section{Experimental Details}
For our experiments, we grew single crystals of Sr$_2$IrO$_4$ using established SrCl$_2$-flux techniques\cite{kimscience}. The stoichiometry of the resulting Sr-214 crystals was confirmed via energy dispersive spectroscopy (EDS) measurements, and a number of Sr-214 crystals were also ground into a powder and checked via x-ray diffraction in a Bruker D2 phaser system. Within resolution, all x-ray peaks were indexed to the reported tetragonal structure (space group $I4_1/acd$, $a=b=5.48$ $\AA$ and $c=25.8$ $\AA$). Neutron diffraction experiments were performed on a $5$ mg crystal with a resolution-limited mosaic of $0.40^{\circ}$ at the C5 triple-axis spectrometer at the Canadian Neutron Beam Center at Chalk River Laboratories. A fixed final energy $E_F=14.5$ meV setup was used with a vertically focusing pyrolytic graphite (PG) crystal (PG-002) as the monochromator and a flat PG-002 analyzer. Collimations were $33^{\prime} -48^{\prime} -51^{\prime} -144^{\prime}$ before the monochromator, sample, analyzer, and detector, respectively along with two PG filters after the sample. The single crystal was mounted within the [H, 0, L] scattering plane within a closed-cycle refrigerator. 

\begin{figure}
\includegraphics[scale=.325]{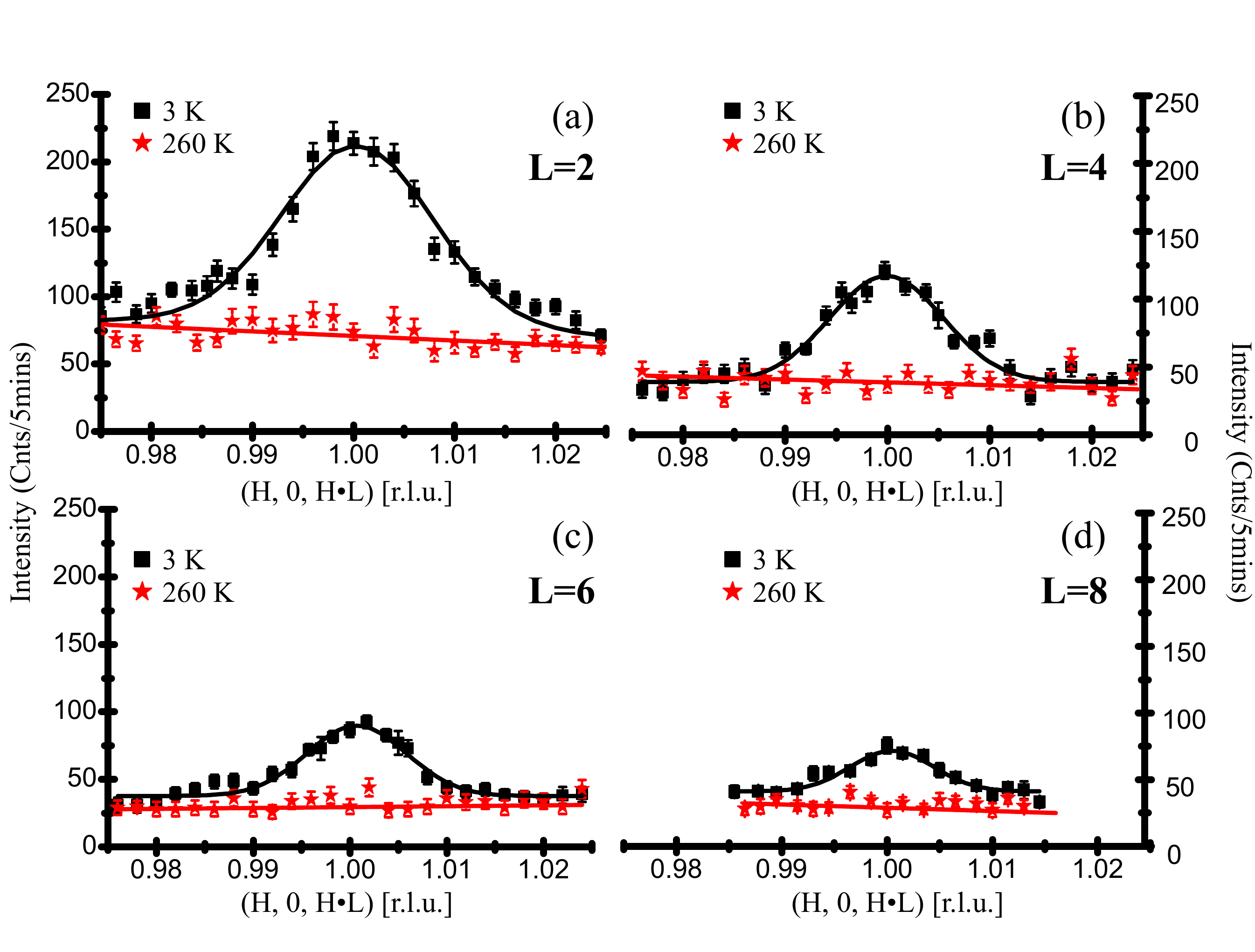}
\caption{Radial Q-scans both above and below T$_N$ through magnetic Bragg peaks at (a) \textbf{Q}$=(1,0,2)$ (b) \textbf{Q}$=(1,0,4)$, (c) \textbf{Q}$=(1,0,6)$, and (d) \textbf{Q}$=(1,0,8)$ positions. Solid black lines are Gaussian fits to the data.}
\end{figure}

\begin{figure}
\includegraphics[scale=.35]{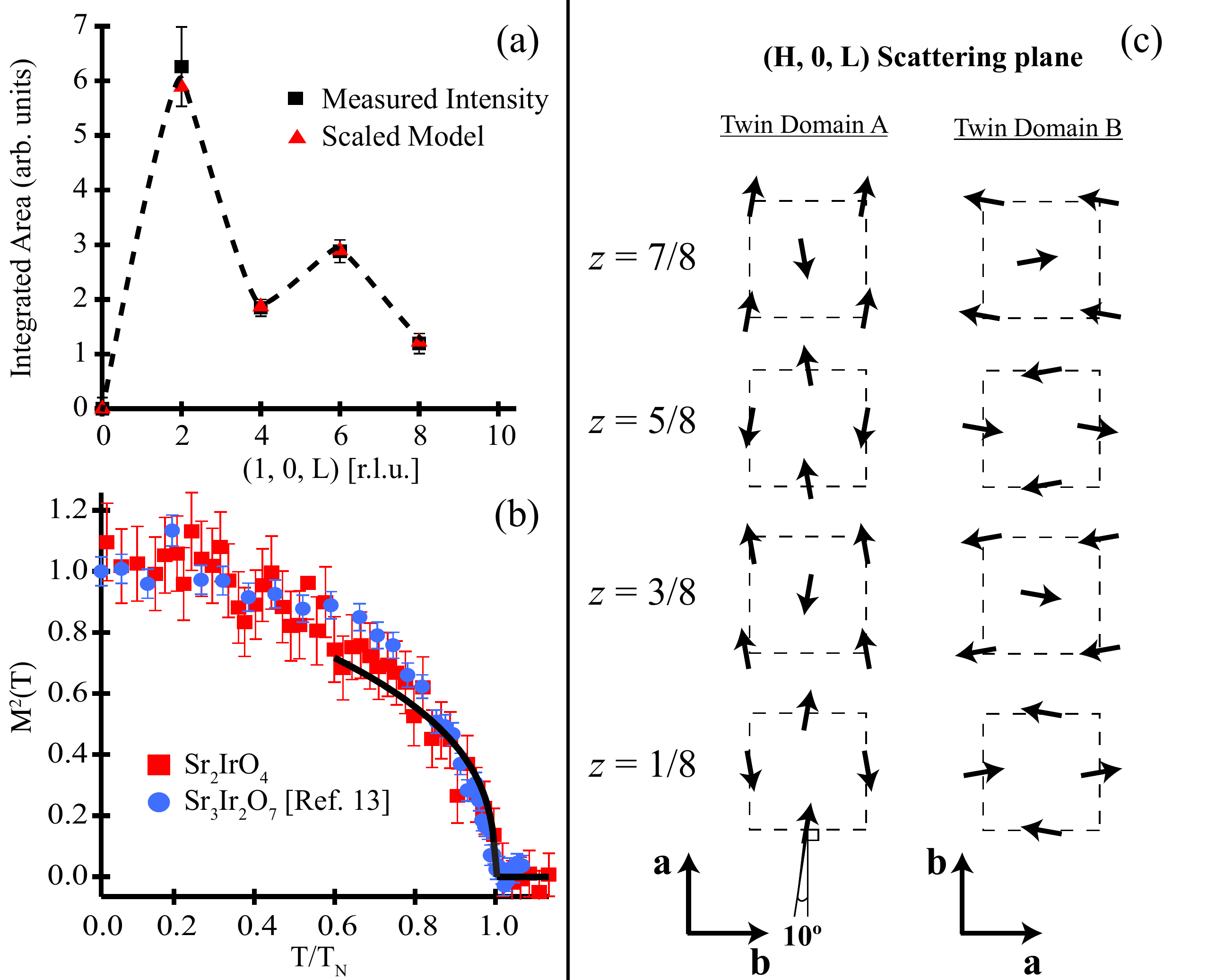}
\caption{(a) Radially integrated intensities (black squares) of magnetic Bragg peaks are plotted as as function of (1,0,L).  Expected intensities from the two domain magnetic model discussed in the text are overplotted as red triangles. (b) Magnetic order parameters squared plotted as a function of reduced temperature for Sr-214 (square symbols) and Sr-327 (circle symbols). (c) In-plane projections of the model of canted AF order utilized in the calculation for Sr-214 moment.  Relative c-axis locations of each plane within the unit cell are denoted to the left of each corresponding plane.}
\end{figure}

\section{Neutron scattering results}
Plotted in Fig. 1 are the results of neutron scattering measurements through the magnetic wave vectors of Sr-214. Specifically, radial scans through the \textbf{Q}=$(1, 0, L)$ positions show that for $L=$ even, magnetic reflections appear below $230$ K; consistent with the known T$_N$ of this system \cite{crawford, cao214}. The correlated order is three dimensional (H- and L-scans not shown) with a minimum in-plane spin-spin correlation length of $\xi=130\pm8$ $\AA$ calculated using the relation $\xi=\sqrt{2ln(2)}w^{-1}$  ($w$ is Gaussian width of the peak in [$\AA^{-1}$]). Given the magnetic structure determined via previous x-ray measurements,\cite{kimscience} the appearance of both $L=4N$ and $L=4N+2$ reflections in the same $[H, 0, L]$ scattering plane implies the presence of two magnetic domains.  As we will argue later, the explanation for these domains is an inherent crystallographic twinning where both the $[1, 0, L]$ and $[0, 1, L]$ structural domains are present within the same experimental scattering plane with moments pointed along an unique in-plane axis.  

Now turning to the radially integrated intensities of the (1,0,L) magnetic peaks, the result of a simple model assuming the previously proposed spin structure of Sr-214\cite{kimscience} (illustrated in Fig. 2(c)) with two crystallographically twinned magnetic domains is overplotted with the experimentally observed intensities in Fig. 2 (a).  This twin-domain model with equal domain populations\cite{momentdirection} agrees remarkably well with the observed neutron intensities, lending support to the assumption of an inherent orthorhombicity/twinning to the spin structure.   The scale factor generated by using this model and normalizing to $(00L)$-type nuclear reflections gives an ordered moment of $\mu_{214}=0.36\pm0.06$ $\mu_B$.  As a direct comparison, if we normalize the moment of Sr-327 from our earlier measurements \cite{dhital} using the same procedure (i.e. same Q-positions) as the current Sr-214 study, the ordered moment for Sr-327 is a nearly identical $\mu_{327}=0.35\pm0.06$ $\mu_B$.    

The magnetic order parameter squared $M^{2}(T/T_N)$ of Sr-214 is overplotted with the known $M^{2}(T/T_N)$ of Sr-327 in Fig. 2 (b).  Both order parameters track one another and power law fits of the form $M^{2}(T)=(1-T/T_N)^{2\beta}$ over the range $0.6\leq T/T_N\leq 1$ in each system yield an identical (within error) $\beta=0.18\pm0.02$ for Sr-214 and $\beta=0.20\pm0.02$ for Sr-327. This fact combined with the comparable moment sizes of each material suggests that each magnetic phase arises from similar interaction symmetries and correlation strengths, and, more importantly, this reinforces that the bulk magnetization data showing a near vanishing moment in Sr-327 \cite{cao327} is simply a reflection of the $c$-axis orientation of the moments in this system (ie. an orientation-based near quenching of its net canted moment). 

\begin{figure}
\includegraphics[scale=.35]{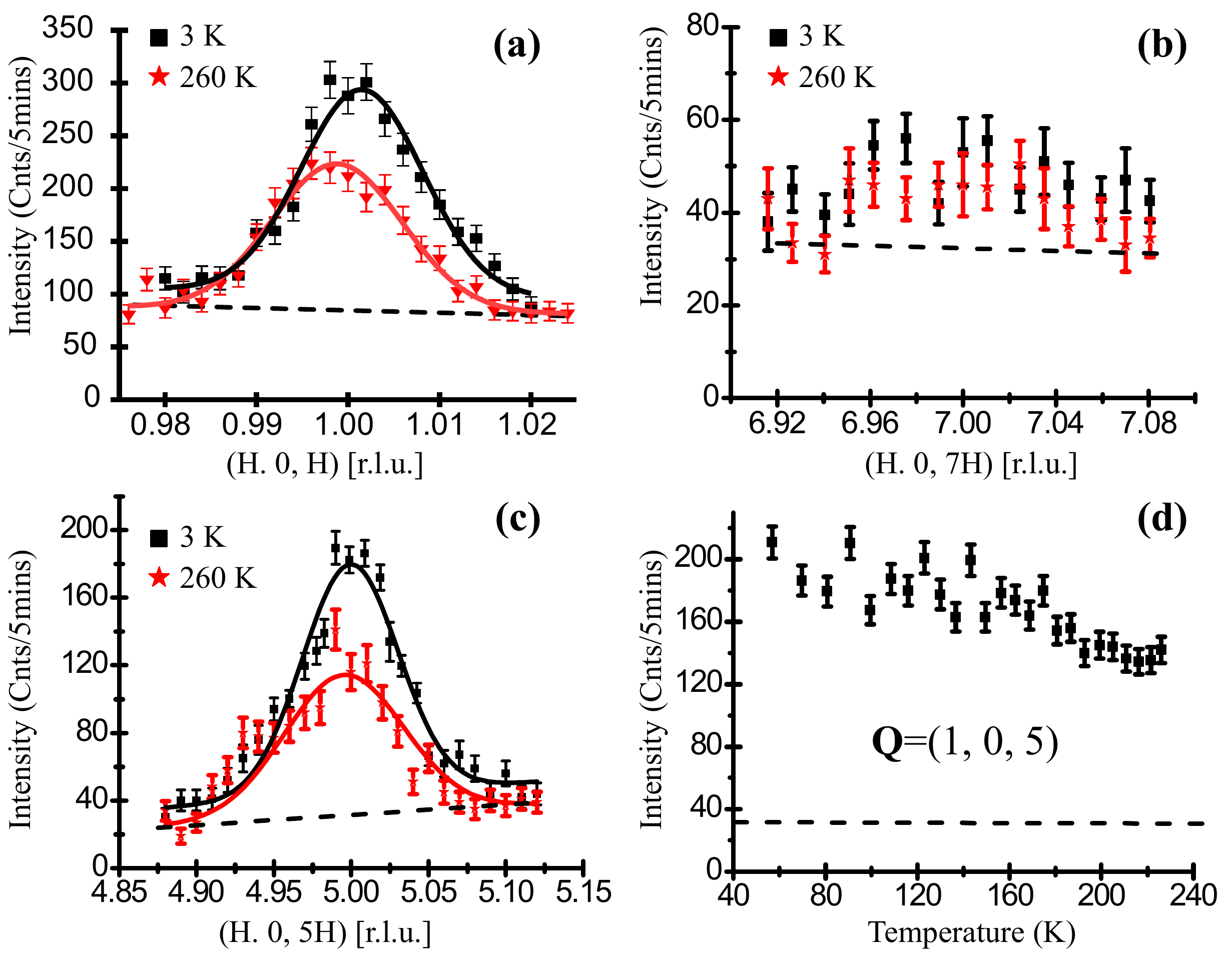}
\caption{Radial scans at 3 K and 260 K through potential nuclear superlattice peaks violating the $I4_1/acd$ space group at (a) \textbf{Q}=(1,0,1), (b) \textbf{Q}=(1,0,5), and (c) \textbf{Q}=(1,0,7). (d) Intensity of the (1,0,5) superlattice peak as a function of temperature.  Dashed line shows the background level for the (1,0,5) peak.}
\end{figure}

The model of twin magnetic domains (domains A and B) in Sr-214 generates magnetic reflections at \textbf{Q$_A$}$=(1, 0, 4N)$ and \textbf{Q$_B$}$=(0, 1, 4N+2)$---apparent as $(1, 0, 4N+2)$ in our data. Its agreement with the data can be seen via the double-peak structure in the intensities plotted in Fig. 2(b). Here the neutron's sensitivity to the moment direction via the $\vec{Q}\times\vec{\mu}\times\vec{Q}$ orientation factor in the scattering cross section generates an enhancement in the $(0, 1, 4N+2)$-type peaks in domain B due to the moments in that domain pointing out of the scattering plane.  As we will discuss later, this suggests that the domain averaging one would expect from a four-domain picture\cite{supplemental} where moments can point along both a- and b-axes is not present and that, instead, moments instead lock to a unique, in-plane axis in a lower crystallographic setting than $I4_1/acd$.    

Supporting the notion of lower structural symmetry, our neutron scattering measurements also reveal the presence of weak superlattice reflections which violate the nominally tetragonal $I4_1/acd$ space group \cite{crawford}. Fig. 3 shows the results of radial scans through a series of $(1,0,L=odd)$ positions forbidden by the $I4_1/acd$ space group.  The intensity of these superlattice peaks continues to decrease with increasing temperature, signifying the presence of a higher temperature lattice distortion similar to the case of Sr-327\cite{dhital}.  While the precise onset temperature of this distortion is unclear, at a minimum, the presence of these forbidden peaks implies a lower structural symmetry than that commonly reported and suggests the notion of magnetic domains arising via crystallographic twinning. A simple survey of likely orthorhombic subgroups of $I4_1/acd$ suggests the candidate space groups of I2$_1$2$_1$2$_1$ or Pnn2; however several tetragonal subgroups are also consistent with these reflections. Further experiments are clearly required to fully understand the final lattice symmetry and to ultimately discern the role of potential oxygen vacancy/defects in the resulting structure and symmetry breaking.   

As a further check of magnetic order in Sr-214, we searched for the presence of $(0,0,L)$ peaks arising from the correlated in-plane canting sequence modulated along the $c$-axis.  Q-scans in Fig. 4 through the magnetic peak position \textbf{Q}$=(0,0,3)$ show the presence of two components of scattering along $L$.  The first is a weak magnetic peak centered at (0,0,3) due to the c-axis modulated in-plane canting orientations in Sr-214.  The second component however is a broad short-range scattering signal that is uncorrelated along $L$ and short-range along $H$.  The in-plane correlation length of this underlying broad lineshape is only $\xi=4\pm1$ $\AA$, reflective of in-plane correlations spanning the length of one unit cell and completely disordered out of the IrO$_2$-plane.  

\begin{figure}
\includegraphics[scale=.5]{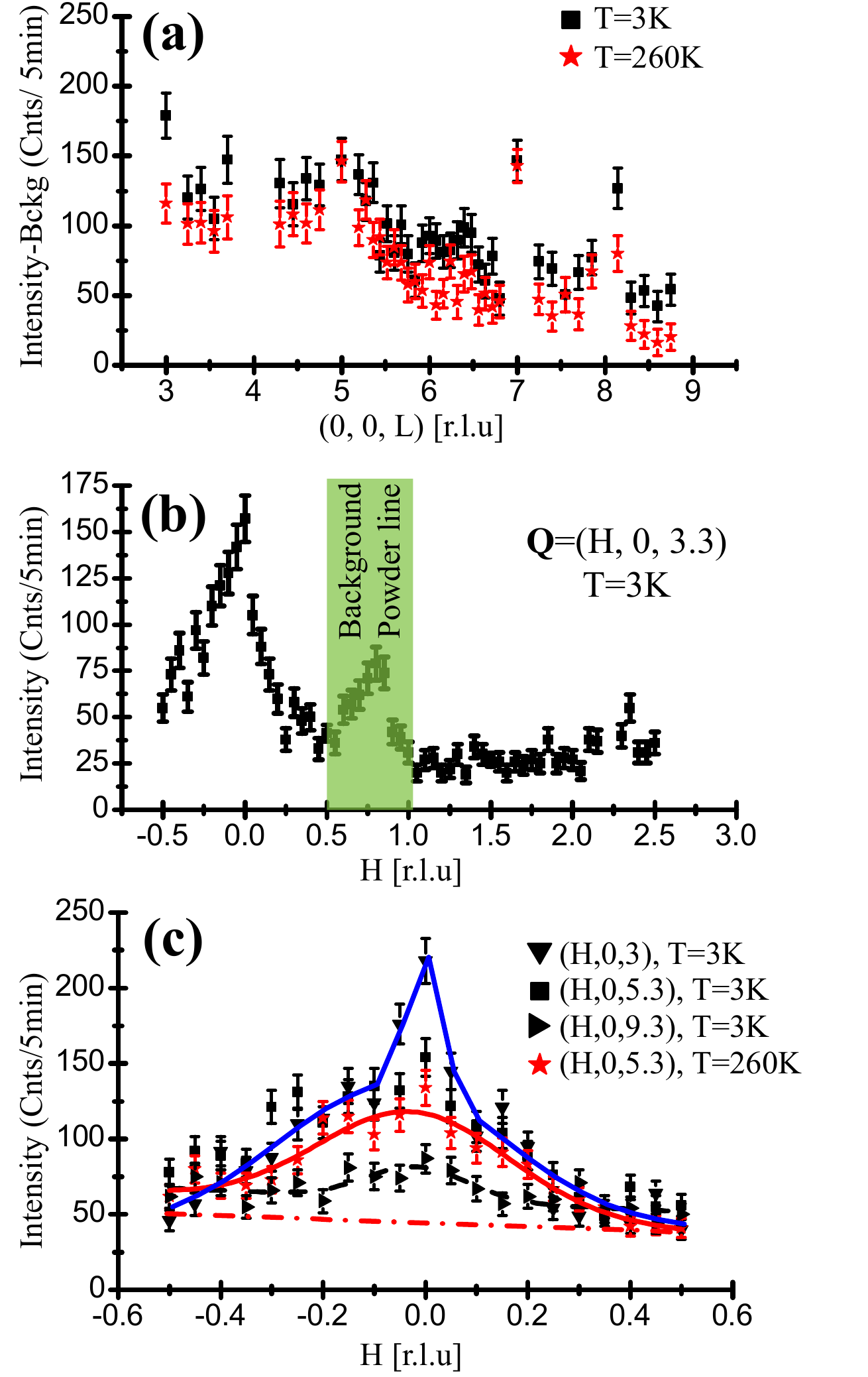}
\caption{(a) Q-scan along the (0,0,L) direction showing a diffuse rod of scattering at both 3 K and 260 K.  Peaks at (0,0,4) and (0,0,8) are primary Bragg peaks. (b) H-scan through the (0,0,3.3) position.  Green box masks background powder scattering from the sample mount. (c) H-scans through the L=3, 5.3, and 9.3 positions.  Dashed line denotes the fit background, and solid lines are Gaussian fits to the data. For the (H, 0, 3) peak, two Gaussian lineshapes were fit to the data.}
\end{figure}

The intensity of this diffuse rod of scattering drops rapidly with increasing $L$ as shown in Fig. 4 (a).  Here the background at \textbf{Q}$=(0.6, 0, L)$ has been subtracted from the diffuse signal, and the signal vanishing is demonstrated explicitly by $H$-scans at increasing $L$-values plotted in Fig. 4 (c). Furthermore, an extended $H$-scan at the non integer \textbf{Q}$=(0,0,3.3)$ position (Fig. 4(b)) shows that the diffuse rod of scattering only extends from the $H=0$ position and is absent at the next allowed nuclear zone center at $H=2$.  This type of rapid drop in intensity curiously tracks the naive expectation of the isotropic Ir$^{4+}$ form factor; however the scattering remains only weakly temperature dependent.  

The precise origin of this short-range scattering is presently unclear; however it likely reflects local disorder in the in-plane rotations of the oxygen octahedra.  The larger relative spectral weight of the diffuse component and the absence of appreciable temperature dependence suggest a structural origin.  Given the rather delicate pattern of out-of-plane rotations of oxygen octahedra comprising the chemical unit cell within the $I4_1/acd$ space group, some degree of local oxygen disorder is likely present, and hints of a similar diffuse component in the nuclear scattering profile where reported by the original polycrystalline refinements of Crawford et al. \cite{crawford}. The degree of oxygen disorder may vary between samples and ultimately account for the variability observed in the transport properties in single crystals of this Sr-214 system \cite{cao214,korneta}.

\section{Discussion}
Structural scattering violating the $I4_1/acd$ space group combined with the twinning observed in the magnetism of this material demonstrates a lower structural symmetry for Sr-214.  At present, a full neutron scattering-based single crystal refinement sensitive to the oxygen distortions has not been performed; however the temperature evolution of this order parameter, at a minimum, demonstrates the presence of an additional structural order parameter in the ground state of this material, similar to the case of Sr-327 \cite{dhital}.  This higher temperature structural distortion likely plays a role in the formation of the electronic gap known to be present above $T_N$ in this material \cite{moon}.  Due to the remaining uncertainty in the crystal structure, a full representational analysis of the magnetic basis vectors allowed within each irreducible representation of the magnetic phase is not currently possible. Despite this limitation however, the absolute ordered moment can still be determined and the presence of magnetic domains can be leveraged to gain insight into the magnetic system.      

First, the total moment we determine agrees within error with the bounds reported in a recently reported powder diffraction measurement\cite{chapon} with $\mu_{polycrystal}=0.29\pm0.04$ $\mu_B$; albeit the moment determined in our single crystal measurement is near the upper bound of the moment derived from the polycrystalline measurement of Lovesey et al.  Our finding that the ordered AF moment of Sr-214 is $\mu_{214}=0.36\pm0.06$ $\mu_B$ is also empirically consistent with the known saturated ferromagnetic moment of $M_{bulk}\approx0.075$ $\mu_B$ \cite{kimscience} that arises from the in-plane canting of all Sr-214 layers.  Using the measured AF moment value and assuming a $10^{\circ}$  basal plane rotation of the oxygen octahedra, the expected canted ferromagnetic moment from our data is $M_{Bulk}=0.063$ $\mu_B$.  This value, although slightly smaller, agrees within error with earlier reports \cite{kimscience, cao214}; however subtle oxygen stoichiometry/disorder effects can also renormalize the AF moment resulting in some sample dependence \cite{kini, cao214}.  The moment determined in our study is also in agreement with early LDA theoretical estimates\cite{yu}. 

The near identical ordered moments of Sr-214 ($\mu_{214}=0.36$ $\mu_B$) and Sr-327 ($\mu_{327}=0.35$ $\mu_B$) reflect the comparable correlation strengths between the two systems.  This result stands in stark contrast to recent RXS studies which claim a substantially smaller moment within Sr-327 relative to Sr-214 \cite{fujiyama}.  Neutron scattering, however, provides a more quantitative assessment of the ordered moment size relative to the second-order RXS process, where recent work has theorized that the RXS signal is potentially comprised of only the orbital component of the total angular momentum \cite{lovesey}.  

The intensity variation of magnetic scattering at the (1, 0, L) positions plotted in Fig. 2 (a) surprisingly suggests that between magnetic domains, moments point along the same unique in-plane axis plus a small degree of canting.  Our two-domain picture of domain formation is presently modeled as stemming from crystallographic twinning and we have assumed the same relative orientations between Ir-site spins in each domain.  If we maintain this assumption of fixed relative spin orientations in all domains and now consider a more general scenario with tetragonal symmetry, a four-domain picture with crystallographically twinned domains whose moments point along the $a$-axis in two of the domains and along the $b$-axis in the other two can be envisioned.   However, due to the neutron orientation factor, in Sr-214 the domain averaged intensities generated at each magnetic reflection differ between these two pictures of domain creation--most notably through the absence of the (1, 0,0) magnetic reflection\cite{supplemental}---and only the two-domain scenario with moments oriented along one unique in-plane direction matches the observed data in Fig. 2 (b). This further suggests the notion of an underlying orthorhombic structural symmetry or, alternatvely, more exotic means of symmetry breaking that bias the in-plane moment orientation such as single-ion anisotropy effects or an in-plane electronic nematicity.

We note here that in our four-domain models of magnetism in this system, we assumed that the orientations of the in-plane moments can be uniformly rotated from a predominantly $a$-axis orientation to a $b$-axis orientation using basis vectors within a single irreducible representation (IR) of the magnetic order.  This is \emph{not} allowed using the IRs resulting from the decomposition of the magnetic phase using the $I4_1/acd$ structural symmetry as shown by Lovesey et al.\cite{chapon}. However, given that our present measurements show that the true crystal structure possesses a lower symmetry, such a uniform rotation does become possible in lower candidate symmetries\cite{supplemental}.  It is also worth noting that if we abandon the assumption that each domain maintains the same relative spin orientations, more exotic four-domain scenarios are possible in which domains with moments oriented along the b-axis also possess a simultaneous phase shift where Ir-moments in the z=1/8 and z=5/8 planes rotate by 180 degrees.  Differentiating the possibility of this scenario from our two-domain picture will require future polarized neutron experiments.         
       
\section{Conclusions}
Our results demonstrate that the ordered magnetic moment in Sr-214 is identical to that observed in its bilayer RP analog, Sr-327.  Scattering from forbidden nuclear reflections and the presence of magnetic twin domains suggest an intrinsic orthorhombicity to the spin structure where oxygen distortions break the $I4_1/acd$ symmetry and evolve as a function of temperature.  Our results demonstrate that, aside from the ordered moment direction, the magnetic and structural phase behaviors in Sr-214 and Sr-327 parallel one another, and suggest that correlation physics likely plays a similar role in each material's respective ground state.  
 
\textsl{Note added in proof:} A related work was posted exploring the structural and magnetic phase behaviors in Sr$_2$IrO$_4$ \cite{fengye}.  Their observations of a high temperature structural distortion in this system and magnetic domain formation are consistent with the results presented here.  

\acknowledgments{
The work at BC was supported by NSF Award DMR-1056625 (S.D.W). SDW acknowledges helpful discussions with Ziqiang Wang, Stephan Rosenkranz, Jeff Lynn, Feng Ye, and J. Fernandez-Baca.}


\begin{thebibliography}{}
\bibitem{momentdirection}  The choice of the a-xis here is arbitrary and equivalent to choosing the moments along the b-axis.
\end{thebibliography}

\end{document}